\documentclass[twocolumn,,showpacs,amssymb,amsmath,aps,pra]{revtex4-1}
\usepackage{graphics}
\usepackage{bm}
\usepackage{amsmath}
\usepackage{amssymb}
\begin{document}

\title{Evaluation of pairwise entanglement in translationally
       invariant systems with the random phase approximation}
\author{J.M.\ Matera, R.\ Rossignoli, N.\ Canosa}
\affiliation{Departamento de F\'{\i}sica-IFLP,
Universidad Nacional de La Plata, C.C. 67, La Plata (1900), Argentina}
\date{\today}

\begin{abstract}	
We discuss a general mean field plus random phase approximation (RPA) for
describing composite systems at zero and finite temperature. We analyze in
particular its implementation in finite systems invariant under translations,
where for uniform mean fields it requires just the solution of simple
local-type RPA equations. As test and application, we use the method for
evaluating the entanglement between two spins in cyclic spin 1/2 chains with
both long and short range anisotropic $XY$-type couplings in a uniform
transverse magnetic field. The approach is shown to provide an accurate
analytic description of the concurrence for strong fields, for any coupling
range, pair separation or chain size, where it predicts an entanglement range
which can be at most twice that of interaction. It also correctly predicts the
existence of a separability field together with full entanglement range in its
vicinity. The general accuracy of the approach improves as the range of the
interaction increases.
\end{abstract}
\maketitle

\section{Introduction}
The random phase approximation (RPA) \cite{BP.53,RS.80,KLT.83} is a well-known
technique in many-body physics. It can be considered as the next step after the
mean field approximation (MFA), being able to describe in a rather simple way
some of the effects induced by the residual interaction, such as collective
excitations \cite{RS.80}. In this work we want to examine its application to
the problem of evaluating pairwise type entanglement in general composite
systems invariant under translations, such as cyclic spin chains with long or
short range couplings in a uniform magnetic field, at both zero and finite
temperature. The fundamental importance of quantum entanglement in different
areas of physics is well recognized, constituting an essential resource for
quantum information science \cite{NC.00,BD.00} and providing a deeper
understanding of quantum correlations in many-body and condensed matter physics
\cite{ON.02,V.03,AOFV.08}. Nonetheless, the evaluation or even the estimation
of entanglement in interacting many-body systems is in general not an easy
task, particularly for long range couplings and finite temperatures, lying
beyond the scope of basic methods like the MFA which rely on separable trial
states.

Here we will show that the MF+RPA can provide a simple general method for
estimating pairwise entanglement, with a complexity which does not exceed that
of solving a {\it local} MF+RPA problem in the case of translationally
invariant systems with uniform mean fields. Its accuracy actually increases for
long range interactions or high connectivity, i.e. for situations where
numerical techniques for evaluating ground states of spin chains (like Quantum
Monte-Carlo \cite{S.97}, DMRG \cite{SW.05} and methods based on matrix product
states \cite{VC.06}) become normally more complex to apply or less accurate. In
any case it allows for a rapid estimation of the main features and their
behavior with the control parameters, leaving the application of more accurate
approaches for a second step. We have previously shown that for fully and
symmetrically connected spin systems (Lipkin-type models \cite{LMG.65,V.06}), a
MF+RPA treatment is indeed able to describe the pairwise entanglement at both
zero and finite temperatures \cite{CMR.07}, becoming exact in the thermodynamic
limit. Its ability to reproduce pairwise entanglement in more general systems
was, however, not examined.

We will first briefly revisit the general MF+RPA formalism derived from the
path-integral representation of the partition function, discussing its
implementation in composite systems and in particular in those which are
translationally invariant.  We next apply the method to finite cyclic spin
$1/2$ chains with general range anisotropic $XYZ$ type couplings. Comparison
with numerical exact results is made for finite chains with anisotropic $XY$
interactions of distinct ranges. The method is able to capture most essential
features of the entanglement between two arbitrary spins away from MF critical
regions, becoming accurate for strong magnetic fields, where it provides an
analytic description of the concurrence. At weak fields the agreement with
exact results is less accurate but improves as the interaction range increases,
being as well able to predict the appearance of a factorizing field
\cite{K.82,A.06,RCM.08} and an infinite entanglement range in its vicinity.

\section{Formalism}
\subsection{General Mean Field+RPA treatment}
We consider a general system of $n$ distinguishable constituents with Hilbert
space dimensions $d_i$, interacting through a general quadratic Hamiltonian
\begin{eqnarray}
 H&=& b^{\mu}O_{\mu}-{\textstyle\frac{1}{2}} V^{\mu \nu }O_{\mu }O_{\nu }\,,
 \label{H}
\end{eqnarray}
where we have adopted tensor sum convention for repeated labels and $O_{\mu}$
stand for general independent linear combinations of {\it local} operators,
i.e., $O_\mu=\sum_{i=1}^n O_{\mu i}$, with $O_{\mu i}=I_1\otimes\ldots
I_{i-1}\otimes o_{\mu i}\otimes I_{i+1}\ldots\otimes I_n$ ($[O_{\mu i},O_{\nu
j}]=0$ if $i\neq j$). We will assume $V^{\mu\nu}=V^{\nu\mu}$, as commutators
$[O_\mu,O_\nu]$ are again linear combinations of local operators and can be
included in the linear term in (\ref{H}). A hamiltonian linear in $O_{\mu}$
represents obviously a non interacting system, being diagonal in a basis of
separable states and requiring just $d_i\times d_i$ local diagonalizations,
whereas $H$ demands in principle a $\prod_i d_i\times \prod_i d_i$
diagonalization, its eigenstates being entangled in general.

The partition function $Z={\rm Tr}\,\exp[-\beta H]$ admits, however, an {\it
exact} representation in terms of linear hamiltonians by means of the auxiliary
field path integral \cite{HS.58}
\begin{eqnarray}
  Z&=&\int{\cal{D}}[\phi]
  {\rm Tr}\,\hat{\cal{T}}
  \exp\left(-\int_0^\beta H[\phi(\tau)]d\tau\right)\,,\label{Zx}\\
  H(\phi)&=&{\textstyle\frac{1}{2}} V^{-1}_{\mu\nu}\phi^{\mu}\phi^{\nu}
  +(b^{\mu}-\phi^{\mu})O_{\mu}\,,\label{hphi}
\end{eqnarray}
where $\phi^{\mu}$ are the auxiliary fields and $\hat{\mathcal T}$ denotes
(imaginary) time ordering. The operator under the trace in Eq.\ (\ref{Zx}) is
just the imaginary time evolution operator associated with the linear
Hamiltonian $H[\phi(\tau)]$, being then a product of {\it local} evolution
operators. Eq.\ (\ref{Zx}) can in principle be evaluated through a Fourier
expansion $\phi^{\mu}(\tau)=\sum_{m=-\infty}^\infty \phi^{\mu}_m e^{i 2\pi
m\tau/\beta}$, with ${\cal D}[\phi]=\prod_m [{\rm det}(\frac{2\pi
V}{\beta})^{-1/2}\prod_\mu d\phi^\mu_m]$.

The MF+RPA treatment, to be abbreviated as CMF \cite{CMR.07},
is obtained by evaluating Eq.\ (\ref{Zx}) in the gaussian
approximation \cite{KLT.83} around the static mean field
$\phi^\mu_m=\delta_{m0}\phi^\mu$ which maximizes
\begin{equation}
Z_{\rm MF}(\phi)={\rm Tr}\exp[-\beta H(\phi)]=e^{-{\textstyle\frac{1}{2}}\beta
V^{-1}_{\mu\nu}\phi^{\mu}\phi^{\nu}}\prod_i z_i(\phi)\,, \label{Zmf}
\end{equation}
where $z_i(\phi)={\rm tr}\exp[-\beta(b^\mu-\phi^\mu)o_{\mu i}]$ is a {\it
local} partition function. It then satisfies the self-consistent equations
\begin{equation}
\phi^{\mu}=V^{\mu\nu}\langle O_{\nu}\rangle_\phi,\;\;
\langle O_{\nu}\rangle_\phi=\beta^{-1}
\sum_i\frac{\partial \ln z_i(\phi)}{\partial\phi^\nu}\,, \label{mf}
\end{equation}
such that $\langle H(\phi)\rangle_\phi=b^{\mu}\langle O_\mu\rangle_\phi-
{\textstyle\frac{1}{2}} V^{\mu\nu}\langle O_\mu\rangle_\phi\langle
O_\nu\rangle_{\phi}$ at a solution. The final result can be written as
 \begin{eqnarray}
  Z_{\rm CMF}&=& Z_{\rm MF}(\phi)C(\phi)\,,\label{Zcmf}\\
 C(\phi)&=&\prod_{m=0}^\infty
 {\rm Det}[\delta^\mu_\nu+V^{\mu\rho}R^m_{\rho\nu}]^{-1
 +\frac{1}{2}\delta_{m0}}\label{C1}\\
 &=&{\rm Det}[\delta^\mu_\nu+V^{\mu\rho}R^0_{\rho\nu}]^{-\frac{1}{2}}
 \prod_{\alpha>0}
 \frac{\omega_\alpha\sinh\frac{\beta\lambda_\alpha}{2}}
{\lambda_\alpha\sinh\frac{\beta\omega_\alpha}{2}}\,,\label{C2}
\end{eqnarray}
where $R^m_{\mu\nu}=\sum_{j=1}^n r^m_{\mu\nu}(j)$ are MF response matrices and
\begin{eqnarray}
 r^m_{\mu\nu}(j)&=&
r_{\mu\nu}(j,\frac{2\pi i m}{\beta})-\delta_{m0}\sum_{\kappa_j}
\langle \kappa_j|o_{\mu j}|\kappa_j\rangle
 \frac{\partial p_{\kappa_j}}{\partial\phi^\nu}\,,\\
r_{\mu\nu}(j,\omega)&=&
\sum_{\kappa_j\neq\kappa'_j}\langle\kappa_j|o_{\mu j}|\kappa'_j\rangle
 \langle\kappa'_j|o_{\nu j}|\kappa_j\rangle\frac{p_{\kappa_j}-p_{\kappa'_j}}
{\varepsilon_{\kappa_j}-\varepsilon_{\kappa'_j}+\omega}\,,\label{Rw}
\end{eqnarray}
are {\it local} responses ($r^0_{\mu\nu}(j)=-\partial\langle o_{\mu
j}\rangle_\phi/\partial \phi^\nu$ is a local susceptibility matrix), with
$(b^\mu-\phi^\mu)o_{\mu j}|\kappa_j\rangle=
\varepsilon_{\kappa_j}|\kappa_j\rangle$, $\langle\kappa_j|\kappa'_j\rangle=
\delta_{\kappa_j\kappa'_j}$ and
$p_{\kappa_j}=e^{-\beta\varepsilon_{\kappa_j}}/z_j(\phi)$. $C(\phi)$ contains
the static ($\delta \phi_{0}^{\mu}$) and quantum ($\delta \phi^{\mu}_{m\neq
0}$) gaussian fluctuations around the mean field and requires just {\it local}
diagonalizations if evaluated through Eq.\ (\ref{C1}). In the closed form
(\ref{C2}), $\lambda_\alpha\equiv
\varepsilon_{\kappa_j}-\varepsilon_{\kappa'_j}$, with $\alpha$ labelling all
pairs of distinct local eigenstates ($\alpha>0$ indicating $\kappa_j>
\kappa'_j$), while $\omega_\alpha$ are the RPA energies, obtained from
\begin{equation}
{\rm Det}[\delta^{\mu}_{\nu}+V^{\mu\rho}R_{\rho\nu}(\omega_\alpha)]=0\,,
\label{wrpa}
 \end{equation}
where $R(\omega)=\sum_j r(j,\omega)$. They are the poles of the RPA response
matrix $[I+R(\omega)V]^{-1}R(\omega)$ and come in pairs of opposite sign.
They can also be obtained as the eigenvalues of the RPA matrix
\begin{equation}
{\cal A}_{\alpha\alpha'}=\lambda_\alpha\delta_{\alpha\alpha'}+
f_\alpha O_{\mu,-\alpha}V^{\mu\nu}O_{\nu\alpha'}\,,\label{Arpa}
\end{equation}
where $f_\alpha=p_{\kappa_j}-p_{\kappa'_j}$, $O_{\mu\alpha}=
\langle \kappa_j|o_{\mu j}|\kappa'_j\rangle$, of dimension
$\sum_j d_j(d_j-1)$.  Let us mention that under a linear transformation
 $O_\mu=U^\nu_\mu \tilde{O}_\nu$, we have $b^\mu\rightarrow
 \tilde{b}^\mu=U^\mu_\nu b^\nu$, $V^{\mu\nu}\rightarrow \tilde{V}^{\mu\nu}=
 U^\mu_{\rho}U^\nu_\lambda V^{\rho\lambda}$,
Eqs.\ (\ref{C1})-(\ref{C2}) being of course independent of the representation.

Eq.\ (\ref{Zcmf}) can be applied away from MF critical points (where the static
determinant in (\ref{C1})-(\ref{C2}) and the lowest RPA energy will vanish, and
where the approach can be improved for $T>0$ by integrating exactly over the
relevant static variables \cite{PBB.91,RCR.98,CMR.07}), becoming accurate for
small $VR$. In the presence of vanishing RPA energies arising due to a mean
field which breaks a continuous symmetry of $H$ \cite{KLT.83}, the product in
(\ref{C2}) remains finite but $\mu,\nu$ in $R^0_{\mu\nu}$ should be restricted
to the {\it intrinsic} static fields, with static orientation variables
integrated out exactly and contributing with a prefactor to (\ref{C2})
\cite{CMR.07}. If $\omega_\alpha\neq 0$ $\forall$ $\alpha$, we may rewrite
(\ref{C2}) as
\begin{equation}
C(\phi)={\rm Det}[\delta^\mu_\nu+V^{\mu\lambda}
{R'}^0_{\lambda\nu}]^{-\frac{1}{2}}
 \prod_{\alpha>0}\frac{\sinh\frac{\beta\lambda_\alpha}{2}}
{\sinh\frac{\beta\omega_\alpha}{2}}\,,\label{C3}
\end{equation}
where ${R'}^0=[R^0-R(0)][1+VR(0)]^{-1}$ vanishes for $T\rightarrow 0$ and the
last factor is just the ratio of partition functions of independent bosons of
energies $\omega_\alpha$ and $\lambda_\alpha$. For $T\rightarrow 0$ the energy
$E_{\rm CMF}=-\frac{\partial\ln Z_{\rm CMF}}{\partial \beta}$ approaches the
usual MF+RPA expression \cite{RS.80} $\langle
H(\phi)\rangle_\phi+{\textstyle\frac{1}{2}} \sum_\alpha
(\omega_\alpha-\lambda_\alpha)$.

For a Hamiltonian representation in terms of purely local operators $O_{\mu
i}$, we should replace $O_\mu$ and $\phi^\mu$ by $O_{\mu i}$ and $\phi^{\mu i}$
in previous expressions,
\begin{equation}
 H = b^{\mu i}O_{\mu i}-{\textstyle\frac{1}{2}}V^{\mu i\nu j}
  O_{\mu i} O_{\nu j}\,,\label{H1}
\end{equation}
and $R^m_{\mu i\nu j}=\delta_{ij}r^m_{\mu\nu}(i)$, such that
$V^{\mu\rho}R^m_{\rho\nu}\rightarrow V^{\mu i\rho j}r^m_{\rho\nu}(j)$. Eqs.\
(\ref{C1})-(\ref{C2}) will then involve in general determinants of matrices
connecting all components. We can assume $V^{\mu i\nu i}=0$, as self-energy
terms are local operators and can in principle be also included in the linear
term.

Although the representation (\ref{H1}) is not necessarily the most convenient
one for evaluating $C(\phi)$, it allows to evaluate two site averages directly
as $\langle O_{\mu i}O_{\nu j}\rangle= 2\beta^{-1}\partial \ln Z/\partial
V^{\mu i\nu j}$, leading in CMF to
\begin{equation}
 \langle O_{\mu i}O_{\nu j}\rangle= \langle O_{\mu i}\rangle_{\phi}\langle
 O_{\nu j}\rangle_{\phi}+\frac{2}{\beta}\frac{\partial\ln C(\phi)}{\partial
 V^{\mu i\nu j}}\,.\label{d}
 \end{equation}
The reduced density matrix for the $i-j$ subsystem can then be recovered by
considering a complete set of local operators. For degenerate symmetry breaking
mean fields Eq.\ (\ref{d}) should be averaged in principle over the different
solutions.

\subsection{Translationally invariant systems}
Let us now consider the case of identical components, i.e., identical Hilbert
spaces ($d_i=d$) and operators ($o_{\mu i}=o_\mu$) at each site, with
$b^{\mu i}=b^{\mu}$ and
\begin{equation}
V^{\mu i\nu j}=v^{\mu\nu}(i-j)\,,
\end{equation}
where $v^{\mu\nu}(n-j)=v^{\mu\nu}(-j)$ for a finite cyclic chain. In this
situation we may conveniently rewrite Eq. (\ref{H1}) as
\begin{equation}
H=n[b^{\mu} \tilde{O}_{\mu 0}- {\textstyle\frac{1}{2}}\sum_{k=0}^{n-1}
\tilde{v}^{\mu\nu}(k)\tilde{O}_{\mu k}\tilde{O}_{\nu,-k}]\,,\label{Hk}
\end{equation}
where $\tilde{v}(k)$ is the (discrete) Fourier transform of $v(j)$,
\begin{equation}
\tilde{v}^{\mu \nu}(k)=\sum_{j=0}^{n-1} e^{-i2\pi k j/n} v^{\mu\nu}(j)\,,
\end{equation}
and similarly, $\tilde{O}_{\mu k}=n^{-1}\sum_{j=1}^{n}e^{i2\pi kj/n}O_{\mu j}$
(such that $O_{\mu j}=\sum_{k} e^{-i2\pi kj/n}\tilde{O}_{\mu k}$). Thus,
$V^{\mu k \nu k'}=n\delta_{k,-k'} \tilde{v}^{\mu \nu}(k)$ in Fourier
representation.

We will also assume a {\it uniform mean field} $\phi^{\mu i}=\phi^\mu$, such
that $\langle O_{\mu i}\rangle_\phi=\langle o_\mu\rangle_\phi$ and hence (Eq.\
\ref{mf}),
\begin{equation}\phi^{\mu}=\tilde{v}^{\mu\nu}(0)\langle o_\nu\rangle_\phi
\,,  \label{mf2}
 \end{equation}
which is an effective {\it local} MF equation depending just on the total
coupling $\tilde{v}^{\mu\nu}(0)=\sum_{j}v^{\mu\nu}(j)$.  Notice that in Fourier
space Eqs.\ (\ref{mf}) become $\phi^{\mu k}=n\tilde{v}^{\mu\nu}(k) \langle
\tilde{O}_{\nu, -k}\rangle_{\phi}$, the uniform solution corresponding to
$\tilde{\phi}^{\mu k}=n\delta^{k0}\phi^\mu$ and leading to $\langle
H(\phi)\rangle_\phi=n[b^\mu\langle o_{\mu}\rangle_\phi -\frac{1}{2}
\tilde{v}^{\mu\nu}(0)\langle o_{\mu}\rangle_\phi\langle o_{\nu}\rangle_\phi]$.

In this case $r_{\mu\nu}^m(i)=r^m_{\mu\nu}$ is site independent, implying
$R^m_{\mu k\nu k'}=n^{-2}\sum_{j}e^{i2\pi (k+k')j/n}r^m_{\mu\nu}(j)=
n^{-1}\delta_{k,-k'}r^m_{\mu\nu}$ and therefore, $(VR^m)^{\mu k}_{\nu
k'}=\delta^k_{k'}\tilde{v}^{\mu\rho}(k)r^m_{\rho\nu}$, {\it diagonal in} $k$.
Hence, Eq.\ (\ref{C2}) becomes
\begin{equation}
C(\phi)=\!\!\prod_{k}\left[{\rm Det}[
\delta^\mu_\nu+\tilde{v}^{\mu\rho}(k)r^0_{\rho\nu}]^{-\frac{1}{2}}
\!\prod_{\alpha>0}\frac{\omega_{\alpha}(k)
\sinh\frac{\beta\lambda_\alpha}{2}}
{\lambda_\alpha\sinh\frac{\beta\omega_{\alpha}(k)}{2}}\right]\,,
 \end{equation}
with $\omega_{\alpha}(k)$ the roots of the {\it local} RPA equation
\[{\rm Det}[\delta^\mu_\nu+\tilde{v}^{\mu\rho}(k) r_{\rho\nu}(\omega)]=0\,,\]
or equivalently, the eigenvalues of the effective {\it local} RPA matrix
$a_{\alpha\alpha'}(k)=\lambda_\alpha\delta_{\alpha\alpha'}+f_\alpha
o_{\mu,-\alpha}\tilde{v}^{\mu\nu}(k) o_{\nu\alpha'},$ of dimension $d(d-1)$.
$C(\phi)$ reduces then to the product of {\it $n$ single site correction
factors with couplings $\tilde{v}^{\mu\nu}(k)$}. These results also hold for
{\it $D$-dimensional} cyclic systems (for instance spins in a torus) provided
$V^{\mu\bm{i}\nu \bm{j}}=v^{\mu\nu}(\bm{i}-\bm{j})$, replacing matrices
$\tilde{v}(k)$ by $\tilde{v}(\bm{k})=\sum_{\bm{j}} e^{-i2\pi \sum_{l=1}^D
k_lj_l/n_l}v(\bm{j})$.

Eq.\ (\ref{d}) will now depend just on the separation $i-j$, becoming
\begin{equation}
\langle O_{\mu i}O_{\nu,i+j}\rangle= \langle
o_{\mu}\rangle_{\phi}^2+\frac{2}{n\beta}\sum_k
 e^{i2\pi kj/n}\frac{\partial\ln C(\phi)}{\partial
 \tilde{v}^{\mu\nu}(k)}\,.\label{d2}
\end{equation}

\section{Application}
\subsection{Finite spin 1/2 chain with general range $XYZ$-type couplings}
We now consider a finite spin $1/2$ cyclic chain in a uniform magnetic field.
The local operators are the spin components $s_{\mu}$, $\mu=x,y,z$, and we will
assume $v^{\mu\nu}(j)=\delta^{\mu\nu} v^\mu(j)$ ($j$-independent principal
axes), with the magnetic field parallel to one of these axes ($z$-axis). This
wide class of systems comprises well-known models such as the Ising and 1-D
$XY$ models with nearest neighbor couplings \cite{LSM.61,ON.02}, as well as the
Lipkin model \cite{RS.80,LMG.65,V.06,CMR.07}, where every pair is identically
coupled. The Hamiltonian reads
\begin{equation}
H=b\sum_i S_{zi}-\sum_{\mu,i\neq j}v^\mu(i-j)S_{\mu i}S_{\mu j}\,,\label{Hsp}
\end{equation}
where we will assume $v^\mu(j)=v^\mu(n-j)=v^\mu(-j)$. Eq.\ (\ref{Hsp}) always
commutes with the ``$S_z$ parity'' $P_z=\prod_j e^{i\pi (S_{zj}+1/2)}$,
entailing $\langle S_{\mu i}\rangle=0$, $\langle S_{\mu i}S_{zj}\rangle=0$ for
$\mu=x,y$ and $j\neq i$ at any $T>0$.

Eqs.\ (\ref{mf2}) for a uniform MF become here
\begin{equation}
\phi^\mu=\tilde{v}^\mu_0{\textstyle\frac{\phi^\mu-b^\mu}{\lambda}}
\tanh{\textstyle\frac{1}{2}}\beta\lambda\,,
\label{mf3}
\end{equation}
where $\tilde{v}^\mu_k=\sum_j v^\mu(j)e^{-i2\pi k j/n}$,
$\lambda=\sqrt{\sum_\mu(\phi^\mu-b^\mu)^2}$ and $b^\mu=(0,0,b)$. We will focus
on the anisotropic attractive case $\tilde{v}^x_0>|\tilde{v}^y_0|$,
$\tilde{v}_0^z\geq 0$, where the lowest solution corresponds to $\phi^y=0$ and
i) $\phi^x=0$ (normal solution), valid for
$|b|>b_c=\tilde{v}^x_0-\tilde{v}^z_0$ or $T>T_c=|b|/\ln\frac{b_c+|b|}{b_c-|b|}$
if $|b|<b_c$, where $\phi^z=-\tilde{v}^z_0\tanh{\textstyle\frac{1}{2}}
\beta\lambda$, or otherwise ii) $\phi^x=\pm|\phi^x|\neq 0$ ({\it degenerate
parity breaking solution}), where $\lambda=\tilde{v}^x_0
\tanh{\textstyle\frac{1}{2}}\beta\lambda$, $\phi^z=-\tilde{v}^z_0 b/b_c$. The
ensuing CMF treatment involves here just $2\times 2$ diagonalizations with  a
{\it single} RPA energy for each value of $k$:
\begin{equation}
Z_{\rm CMF}=e^{-n\beta\sum_\mu\!\tilde{v}^\mu_0
\langle s_\mu\rangle_\phi^2}
(2\cosh{\textstyle\frac{1}{2}}\beta\lambda)^n\prod_k c^0_k
\frac{\sinh\frac{\beta\lambda}{2}}
{\sinh\frac{\beta\omega_k}{2}}\,,
\end{equation}
where, defining $\gamma_\mu=\frac{\phi^\mu-b^\mu}{\lambda}$ and
$f=\tanh{\textstyle\frac{1}{2}} \beta\lambda$,
\begin{eqnarray}
\omega_k&=&\lambda\sqrt{(1-f\tilde{v}^y_k/\lambda)[1-f(\gamma_z^2\tilde{v}^x_k
+\gamma_x^2\tilde{v}^z_k)/\lambda]}\,,\label{wk}\\
c^0_k&=&\frac{1}{\sqrt{1-{\textstyle\frac{1}{2}}
\beta(1-f^2)\frac{\lambda(\gamma_z^2\tilde{v}^z_k+ \gamma_x^2\tilde{v}^x_k)-f
\tilde{v}^x_k\tilde{v}^z_k}
{\lambda-f(\gamma_z^2\tilde{v}^x_k+\gamma_x^2\tilde{v}^z_k)}}}\,.
\end{eqnarray}

The spin correlation $\alpha_{\mu j}\equiv \langle S_{\mu i}S_{\mu,i+j}\rangle$
can be evaluated as
\begin{equation}
\alpha_{\mu j}=\frac{1}{n\beta}\frac{\partial\ln Z_{\rm CMF}}{\partial
v^\mu(j)}=\langle s_\mu\rangle_\phi^2+\alpha^c_{\mu j}\,,
\label{al}
\end{equation}
where $\langle s_\mu\rangle_\phi={\textstyle\frac{1}{2}}\gamma_\mu f$ and
$\alpha_{\mu j}^c=\frac{1}{n\beta} \sum_k e^{i2\pi kj/n}\frac{\partial\ln
C(\phi)}{\partial \tilde{v}^\mu_k}$. The reduced two-site density matrix in the
standard basis can then be recovered as
\[\rho_{i,i+j}=
\left(\begin{array}{cccc}p^+_{j}&0&0&
\alpha_{xj}-\alpha_{yj}\\0&\frac{1}{4}-\alpha_{zj}
&\alpha_{xj}+\alpha_{yj}&0\\0&\alpha_{xj}+\alpha_{yj}&
\frac{1}{4}-\alpha_{zj}&0\\
 \alpha_{xj}-\alpha_{yj}&0&0&p^-_{j}\end{array}\right),\]
where $p^\pm_j=\frac{1}{4}+\alpha_{zj}\pm\langle s_z\rangle$ and $\langle
s_z\rangle= -\beta^{-1}\partial \ln Z/\partial b$.

We are here interested on the {\it pairwise concurrence} $C_j$ \cite{W.98}, a
measure of the entanglement between spins $i$ and $i+j$, given in this system
by $C_j={\rm Max}[C^+_j,C^-_j,0]$, where
\begin{eqnarray}
C^+_j&=& 2[|\alpha_{xj}-\alpha_{yj}|+\alpha_{zj}-1/4]\,,\label{cp}\\
C^-_j&=& 2[|\alpha_{xj}+\alpha_{yj}|-\sqrt{(1/4+\alpha_{zj})^2-\langle
 s_z\rangle^2}]\,, \label{cm}
 \end{eqnarray}
represent a parallel or antiparallel concurrence respectively \cite{A.06}. Just
one of $C^{\pm}_j$ can be positive in a given state. Note that $C^{\pm}_j\leq
0$ at the MF level ($\alpha_{\mu j}=\langle s_\mu\rangle_\phi^2$).

\subsection{Concurrence for strong fields}
Let us first examine the concurrence for strong fields $b\gg b_c$, where the
mean field solution is always normal (and uniform) and the ground state is the
fully aligned state plus small corrections.  Eq.\ (\ref{wk}) becomes
\begin{equation}{\textstyle\omega_k=
\lambda[1-f\frac{\tilde{v}^+_k}{\lambda}-{\textstyle\frac{1}{2}}
f^2\frac{(\tilde{v}^-_k)^2}{\lambda^2}- {\textstyle\frac{1}{2}}
f^3\frac{(\tilde{v}_k^-)^{2}\tilde{v}^+_k}{\lambda^3}+
O(f\frac{v}{\lambda})^4]}\label{wk2}\end{equation}
where $\lambda=b+f v_z$,
$\tilde{v}^{\pm}_k={\textstyle\frac{1}{2}}(\tilde{v}^x_k\pm\tilde{v}^y_k)$, and
we have assumed $v^\mu(j)=O(v)$. In this regime the exact GS concurrence can
only be parallel. Up to $O(v/\lambda)^2$, Eqs.\ (\ref{al})--(\ref{wk2}) then
lead at $T=0$ to
\begin{equation}
C^+_j\approx\left|\frac{v_-(j)}{\lambda}+
\frac{{\textstyle\sum_{i=1}^{n-1}v_+(j-i)v_-(i)}}{\lambda^2}\right|-
\frac{{\textstyle\sum_{i=1}^{n-1}v_-^2(i)}}{2\lambda^2}\,, \label{Cb}
 \end{equation}
where $v_{\pm}(j)=(v^x(j)\pm v^y(j))/2$. Hence, pairs connected
by $v_-(j)$ will exhibit in this limit a {\it parallel concurrence of first
order} in $v/\lambda$, whereas those unconnected may still exhibit a parallel
concurrence of {\it second order} in $v/\lambda$ if linked by the {\it
convolution} of $v_+$ with $v_-$. This entails that for an anisotropic
interaction of range $L$ ($v_{\pm}(j)=0$ for $j>L$ and $v_{-}(L)\neq 0$) the
$T=0$ entanglement range for $|b|\gg b_c$ {\it can be at most twice the
interaction range}. Comparison with exact perturbation theory indicates that
for high fields, Eq.\ (\ref{Cb}) is actually {\it exact} for any $n$ but up to
the {\it first non-zero} order. For instance, in the nearest neighbor $XY$ case
$v^\mu(j)=v^\mu(\delta_{j1}+\delta_{j,n-1})/2$, with $v^z=0$,  Eq.\ (\ref{Cb})
leads to $C_j^+=0$ if $j\geq 2$ and
\begin{equation}
C^+_1\approx \frac{|v_-|}{2b}\,,\;\;
 C^+_2\approx \frac{|v_-|(|v_+|-|v_-|)}{4 b^2}\,, \label{C12}
 \end{equation}
with $v_{\pm}=(v^x\pm v^y)/2$, which coincide, up to $O(v/b)$ and $O(v/b)^2$
respectively, with the exact result for the concurrence obtained with the
Jordan-Wigner transformation. Hence, in this limit there will be $O(v/b)^2$
concurrence between second neighbors if $|v_+|>|v_-|$.

For $T>0$, the main thermal corrections to (\ref{Cb}) will arise from the
decrease of the MF contribution $\langle s_\mu\rangle^2_\phi$ to $\alpha_{\mu
j}$, leading to
\begin{equation}
C_j^+(T)\approx C^+_j(0)-2e^{-\beta \lambda}\,,
\label{Cb2}
\end{equation}
for sufficiently low temperatures such that $C^+_j(T)\geq 0$, where
$\lambda=b+v_z$ and $C^+_j(0)$ is the $T=0$ value (\ref{Cb}).  We have
neglected in (\ref{Cb2}) thermal corrections to $\alpha_{\mu j}^c$, which will
lead to higher order terms in $v/\lambda$. From (\ref{Cb2}) we may estimate
the limit temperature for pairwise concurrence at high fields,
\begin{equation}
T^+_j\approx \lambda/\ln[2/C^+_{j}(0)]\,,\label{TL}\end{equation}
which will {\it increase} almost linearly with increasing $b$
($T_j^+\approx O(\frac{b}{\ln(b/v)}))$.

\subsection{Separability field}
Let us now assume a common range such that $v^\mu(j)=r(j)v^\mu$, with $\sum_j
r(j)=1$ ($\tilde{v}^\mu_0=v^\mu$). Anisotropic chains with $v^z<v^y<v^x$ and
$r(j)\geq 0$ will exhibit a factorizing field \cite{K.82,A.06,RCM.08}
$b_s=\sqrt{(v^x-v^z)(v^y-v^z)}<b_c$  where the degenerate parity breaking MF
states become {\it exact} ground states  and $C_j$ vanishes for large $n$
\cite{RCM.08}, changing from antiparallel ($|b|<b_s$) to parallel ($|b|>b_s$).
It is verified that at $T=0$ and $b=b_s$, $\alpha_{\mu j}^c=0$ for $j\neq 0$ in
(\ref{al}), entailing $C^{\pm}_j=0$ $\forall j>0$ also in CMF. Expansion of
$\omega_k$ around $b_s$ actually leads at $T=0$ to
\[\omega_k=v^x[1-r_k\frac{v^y}{v^x}+r_k\frac{b_s}{b_c}\frac{b-b_s}{v^x}
 +O(\frac{b-b_s}{v^x})^2]\,,\]
where $r_k=\sum_{j}e^{-i2\pi jk}r(j)$, implying, up to $O(\frac{b-b_s}{v^x})$,
\begin{eqnarray}
 C^{\pm}_j&\approx&\pm \gamma_j~\frac{b_s}{b_c}\frac{b-b_s}{v^x}\,,\\
\gamma_j&=&\frac{1}{n}\sum_{k}\frac{e^{i2\pi k j}r_k}{1-r_k v^y/v^x}=
 \sum_{m=0}^\infty (\frac{v^y}{v^x})^m r^{m+1}(j)\,,\label{gj}
 \end{eqnarray}
where $r^m(j)\equiv\sum_{i}r(j-i)r^{m-1}(i)$ ($m\geq 2$) denotes the $m^{th}$
convolution of $r(j)$. For {\it any} finite coupling range satisfying $r(j)>0$
for $1\leq j\leq L$ and $0$ otherwise,  Eq.\ (\ref{gj}) yields $\gamma_j> 0$
for $j=1,\ldots,n$. Therefore, CMF will predict in this case {\it full
entanglement range in the immediate vicinity of $b_s$}, with $C_j$ changing
from antiparallel to parallel as $b$ crosses $b_s$, which is in agreement with
the general exact result \cite{RCM.08}. The slope of $C^\pm_j(b)$ at $b=b_s$
is, however, not necessarily exact in CMF.

\subsection{Comparison with exact results in finite chains}
\begin{figure}
\centerline{
\scalebox{.8}{\includegraphics{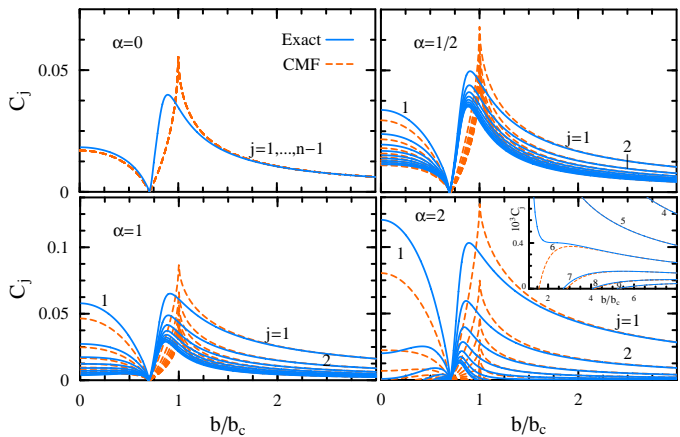}}}

\caption{(Color online) Comparison between exact (solid lines) and CMF (dashed
lines) results for the ground state concurrence $C_j$ of spin pairs with
separation $j$ as a function of the transverse magnetic field $b$, for a long
range $XY$ coupling $v^\mu(i-j)\propto v^\mu/|i-j|^{\alpha}$, with
$\chi=v_y/v_x={\textstyle\frac{1}{2}}$ and $n=18$ spins. The concurrences
vanish at the factorizing field $b_s=\sqrt{\chi}b_c$, where $b_c=v_x$ denotes
the MF critical field. The inset depicts the reentry at high fields of the
concurrence of distant pairs for $\alpha=2$.} \label{f1}
\end{figure}

\begin{figure}
\centerline{\hspace*{-.3cm}\scalebox{.5}{\includegraphics{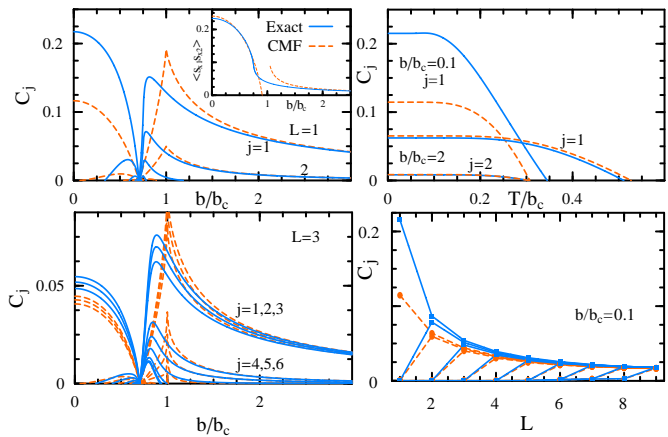}}}

\caption{(Color online) Results for finite range $XY$ couplings at the same
anisotropy $\chi=1/2$. Top panels depict the concurrence for nearest neighbor
coupling ($L=1$) and $n=100$ spins, as a function of the transverse field at
$T=0$ (left) and as a function of temperature at fixed fields (right). The
inset depicts the spin correlation $\langle S_{x1}S_{x2}\rangle$ at $T=0$.
Bottom: Results for interaction range $L$ and constant strength
($v^\mu(i-j)\propto v^\mu$ if $|i-j|\leq L$ and $0$ otherwise) for $n=18$ spins
at $T=0$. The left panel corresponds to $L=3$ (third neighbor coupling) while
the right panel depicts the concurrences $C_j$ as a function of the range $L$
at fixed low field.} \label{f2}
\end{figure}

Illustrative results for anisotropic $XY$ couplings ($v^z(j)=0$) with different
ranges are shown in Figs.\ \ref{f1}-\ref{f2} as a function of the transverse
field. We first consider in Fig.\ \ref{f1} a long range coupling of the form
$v^\mu(j)\propto v^\mu/|j|^\alpha$ for $1\leq |j|\leq n/2$, with
$\tilde{v}^\mu_0=v^\mu\geq 0$, where exact ground state results for $n=18$
spins have been obtained by direct diagonalization. We have selected an
anisotropy $\chi\equiv v^y/v^x=1/2$, in which case the factorizing field is
$b_s=\sqrt{\chi}b_c\approx 0.71 b_c$, with $b_c=v^x$.  As predicted by CMF, at
$b=b_s$ the exact concurrence is seen to vanish for all $\alpha$, reaching
always {\it full range} in its vicinity (for finite $n$ the exact result
actually approaches at $T=0$ exponentially small $\alpha$ and $j$-independent
finite lateral limits  \cite{RCM.08} $C^{\pm}=(1-\chi)\frac{\chi^{(n/2-1)}}{1
\pm\chi^{n/2}}$ at $b=b_s$, with $C^{\pm}\approx 0.002$ for $n=18$ and
$\chi=1/2$, not predicted by CMF).

The $\alpha=0$ case corresponds to the Lipkin model \cite{V.06}, where
$v^\mu(j)=v^\mu/(n-1)$ and $\tilde{v}^\mu_k=v^\mu(n\delta_{k0}-1)/(n-1)$. In
this case $C^{\pm}_j=C^{\pm}$ $\forall$ $j$, with $C^{\pm}<2/n$ \cite{KBI.00}
due to the monogamy property \cite{CKW.00}. CMF is here quite accurate for all
fields values away from $b_c$, providing the exact result for the rescaled
concurrence $nC$ for large $n$ \cite{CMR.07}.

As $\alpha$ increases, CMF remains accurate for high fields $b\agt 1.5 b_c$,
where the concurrence is correctly described by Eq.\ (\ref{Cb}), i.e.,
$C_j\propto (v_-/b)/|j|^{\alpha}$. For sufficiently large $\alpha$ Eq.\
(\ref{Cb}) actually predicts a weak {\it reentry} of the concurrence $C_j$ at
strong fields for large separations $j$, since the last second order term in
(\ref{Cb}) will be negative and greater than the first order term for not too
strong fields if $j$ is sufficiently large.  This reentry is confirmed in the
exact results for large separations, as seen here for $\alpha=2$ (inset of
bottom right panel). CMF looses precision for low fields $|b|\alt b_c$,
although for $\alpha\leq 1$ it is still quite reliable for $|b|<b_s$, where its
accuracy increases as $j$ increases. Notice also that for $\alpha\leq 1$ we
obtain for $n=18$ full range concurrence at {\it all} fields, whereas for
$\alpha=2$ the concurrence becomes very short ranged at low fields ($j\leq 3$),
being non-zero for large $j$ just in the vicinity of $b_s$ or at very strong
fields, i.e., where the nearest neighbor concurrence becomes small, in
agreement with the monogamy property. This behavior is qualitatively reproduced
in CMF. Let us finally mention that for $\alpha=2$, results for the first few
$C_j$ will remain stable as $n$ increases (as $\sum_{j}1/j^\alpha$ is in this
case convergent), those of CMF remaining close to those depicted for $n=18$.

Fig.\ \ref{f2} depicts results for finite range couplings of constant strength,
i.e., $v^\mu(j)=\frac{1}{2}v^\mu/L$ for $|j|\leq L$ and 0 otherwise (such that
$\tilde{v}^\mu_0=v^\mu$), at the same anisotropy. For nearest neighbor
coupling, which corresponds to the $\alpha\rightarrow\infty$ limit of the
previous case, exact results for any finite $n$ and $T$ can be obtained with
the Jordan-Wigner transformation \cite{LSM.61} plus parity projection
\cite{RCM.08}. CMF is again confirmed to be accurate for high fields for
both $j=1$ and $j=2$ (Eq.\ (\ref{C12})), while for $|b|<b_c$ it provides only a
qualitative agreement (with correct predictions like the full entanglement
range in the vicinity of $b_s$),  even though it is still reliable for standard
observables like the spin correlation $\alpha_{x1}$ away from $b_c$ (inset
in the upper left panel). The thermal behavior of $C_j$ is also correctly
described by CMF away from $b_c$, as seen in the upper right panel, where exact
results confirm the increase in the limit temperatures $T_1$ and $T_2$ for high
fields as predicted by Eq.\ (\ref{TL}).

Nevertheless, the accuracy of CMF at low fields improves as soon as the range
$L$ is increased, i.e., as $v^\mu(j)/v^\mu$ decreases. For instance, results at
$b/b_c=0.1$ significantly improve already for $L\geq 2$, as seen in the bottom
right panel, while for $L=3$ CMF is seen to provide the correct general picture
except in the vicinity of $b_c$ (bottom left panel). In particular, the
concurrence range for high fields is seen to be again twice the coupling
range, in agreement with Eq.\ ($\ref{Cb}$) (actually, for $j=6$ both the first
and second order terms in Eq.\ (\ref{Cb}) vanish for $\chi=1/2$ and $L=3$, and
an expansion up to $O(v^x/b)^3$ is required, $C_6(b)$ being still positive in
both CMF and the exact results). The splitting of the concurrences $C_j$ for
$j=1,2,3$ is as well a second order effect.

\section{Conclusions}
We have examined a general MF+RPA treatment for describing composite systems
with quadratic interactions at both zero and finite temperature, showing that
it becomes particularly simple for finite translationally invariant systems
with uniform mean fields. The approach is capable of reproducing the main
features of the pairwise entanglement, for all pair separations, in cyclic spin
1/2 chains with anisotropic $XY$ couplings of different ranges, away from MF
critical regions. It also provides the correct asymptotic behavior of the
concurrence for strong fields, where it predicts interesting features like the
possibility of a reentry of the pairwise concurrence for large separations, as
well as an entanglement range which can be at most twice that of the
interaction for finite range couplings, which were confirmed in the exact
results. It also predicts the factorizing field and the full entanglement range
in its immediate vicinity.

The method is specially suited for treating systems with high connectivity or
long range interactions, where its accuracy improves. Let us remark that the
individual components are in principle arbitrary in the present formalism. They
could be also chosen as small arrays of coupled spins or subsystems treated
exactly, leaving the RPA for the remaining interactions, a possibility which is
currently under investigation and which could improve results for finite range
couplings or dimer type chains. The extension to higher dimensions is as well
straightforward.

J.M.M. and N.C. acknowledge support of CONICET, and R.R. of CIC, of Argentina.


\begin{thebibliography}{999}
\bibitem{BP.53} D.\ Bohm and D.\ Pines, Phys.\  Rev.\ {\bf  92},
               609 (1953).
\bibitem{RS.80} P.\ Ping and P.\ Schuck, {\it The nuclear Many-Body problem}
 Springer  (New York) (1980).
\bibitem{KLT.83} A.K.\ Kerman, S.\ Levit and T.\ Troudet, Ann.\ of Phys.\ (NY)
         {\bf 148}, 436 (1983).
\bibitem{NC.00}M.A.\ Nielsen and I. Chuang, {\it Quantum Computation and
               Quantum Information}, Cambridge Univ. Press (2000).
\bibitem{BD.00}C.H.\ Bennett and D.P.\ DiVincenzo, Nature {\bf 404}, 247
 (2000).
\bibitem{ON.02} T.J.\ Osborne, M.A.\ Nielsen, Phys.\  Rev.\ A {\bf  66},
               032110 (2002).
\bibitem{V.03} G.\ Vidal, J.I.\ Latorre, E.\ Rico, A.\ Kitaev,
              Phys.\ Rev.\ Lett.\ {\bf 90}, 227902 (2003).
\bibitem{AOFV.08} L.\ Amico, R.\ Fazio, A.\ Osterloh and
           V.\ Vedral, Rev.\ Mod.\ Phys.\ {\bf  80}, 517 (2008).
\bibitem{S.97} T.\ Kashiwa, Y.\ Ohnuki, and  M.\ Susuki,  {\it Path Integral
         Methods}, Oxford Univ. Press (1997).
\bibitem{SW.05} U.\, Schollw\"ock, Rev.\ Mod.\ Phys.\ {\bf  77}, 259 (2003).
\bibitem{VC.06}F.\ Verstraete, J.I.\ Cirac, Phys.\ Rev.\ B {\bf 73} 094423
 (2006).
\bibitem{LMG.65} H.\ J.\ Lipkin, N.\ Meshkov, and A.\ J.\ Glick, Nucl.\ Phys.\
{\bf 62}, 188 (1965).
\bibitem{V.06} J.\ Vidal, Phys.\ Rev.\ A {\bf 73} 062318 (2006);
S.\ Dusuel, J.\ Vidal, Phys.\ Rev.\ B {\bf 71}, 224420 (2005).
\bibitem{CMR.07} N. Canosa, J.M.\ Matera, and R. Rossignoli, Phys.\
Rev.\ A {\bf  76} 022310 (2007);
J.M. Matera, R.\ Rossignoli, N.\ Canosa, Phys.\ Rev.\ A
{\bf 78} 012316 (2008).
\bibitem{K.82}
J.\ Kurmann, H.\ Thomas, and G.\ M\"uller, Physica A
 {\bf 112}, 235 (1982).
\bibitem{A.06}
 L.\ Amico  et al, Phys.\ Rev.\ A {\bf 74}, 022322 (2006).
\bibitem{RCM.08}
 R.\ Rossignoli, N.\ Canosa, J.M. Matera,  Phys.\ Rev.\ A
  {\bf 77}, 052322 (2008).
\bibitem{HS.58} R.\ L.\ Stratonovich, Dokl.\ Akad.\ Nauk SSSR {\bf 115},
1097 (1957).  J.\ Hubbard, Phys.\ Rev.\ Lett.\ {\bf 3}, 77 (1959).
\bibitem{PBB.91}
G.\ Puddu, P.F.\ Bortignon, and R. Broglia, Ann.\ Phys.\ (N.Y.)
{\bf 206}, 409 (1991).\\
H.\ Attias, Y. \ Alhassid,  Nucl. Phys.\ A {\bf 625}, 565 (1997).
\bibitem{RCR.98}
R.\ Rossignoli, N.\ Canosa, P. Ring,
                Phys.\ Rev.\ Lett.\ {\bf 80},
                1853 (1998); Ann.\ of Phys.\ (NY) {\bf  275}, 1 (1999).
\bibitem{LSM.61} E.\ Lieb, T.\ Schultz, and D.\ Mattis,
 Ann.\ of Phys.\ (NY) {\bf  16}, 407 (1961).
\bibitem{W.98}
S.\ Hill and  W.K.\ Wootters, Phys.\ Rev.\ Lett.\ {\bf 78}, 5022
 (1997); W.K.\ Wootters, Phys.\ Rev.\ Lett.\ {\bf 80}, 2245 (1998).
\bibitem{KBI.00}
 M.\ Koashi, V.\ Buzek, and N. Imoto,
Phys.\ Rev.\ {\bf A} 62, 050302(R) (2000).
\bibitem{CKW.00}
V.\ Coffman, J.\ Kundu and W.K.\ Wootters,
              Phys.\ Rev.\ A {\bf 61} 052306 (2000);
              T.J.\ Osborne and F. Verstraete,
              Phys.\ Rev.\ Lett.\ {\bf 96}, 220503 (2006).
 \end{thebibliography}
\end{document}